\documentclass[%
 aip,
 amsmath,amssymb,
 reprint,%
]{revtex4-1}

\usepackage{graphicx}% Include figure files
\usepackage{dcolumn}% Align table columns on decimal point
\usepackage{bm}% bold math
\usepackage{color}

\usepackage[utf8]{inputenc}
\usepackage[T1]{fontenc}
\usepackage{mathptmx}
\usepackage{etoolbox}

\makeatletter
\def\@email#1#2{%
 \endgroup
 \patchcmd{\titleblock@produce}
  {\frontmatter@RRAPformat}
  {\frontmatter@RRAPformat{\produce@RRAP{*#1\href{mailto:#2}{#2}}}\frontmatter@RRAPformat}
  {}{}
}%
\makeatother
\begin{document}
%============================================================================
\preprint{AIP/123-QED}
%============================================================================

%\title[Electron acceleration in microwave-driven plasma wakefields]{Electron acceleration in microwave-driven plasma wakefields in rectangular waveguides}

\title[Numerical study of electron acceleration by microwave-driven plasma wakefields in rectangular waveguides.]{Numerical study of electron acceleration by microwave-driven plasma wakefields in rectangular waveguides.
}

%============================================================================
\author{Jesús E. López}
\affiliation{School of Physics, Universidad Industrial de Santander, Bucaramanga, Colombia, 680002}
%----------------------------------------------------------------------------
\author{Eduardo A. Orozco-Ospino}
\email{eaorozco@uis.edu.co}
\affiliation{School of Physics, Universidad Industrial de Santander, Bucaramanga, Colombia, 680002}
%============================================================================
\date{\today}% It is always \today, today,
\begin{abstract}
Plasma-based acceleration schemes have attracted sustained interest as a pathway toward compact particle accelerators, owing to the large electric fields supported by plasmas. Although recent studies have demonstrated the excitation of plasma wakefields using high-power microwave pulses in plasma-filled waveguides, the conditions required for efficient electron acceleration in such configurations remain insufficiently characterized. In this work, we investigate the acceleration of externally injected electrons by microwave-driven plasma wakefields in rectangular waveguides filled with low-density plasma. Three-dimensional particle-in-cell simulations are employed to analyze the dynamics of electron injection and energy gain under both reduced and fully self-consistent numerical models. The results show that electron acceleration is strongly dependent on the injection phase and initial velocity. Optimal acceleration is achieved when electrons are pre-accelerated to velocities close to the group velocity of the driving microwave pulse. For the parameters considered, energy gains of the order of $10^2 \mathrm{keV}$ are obtained over interaction lengths of the order of meters, while maintaining a quasi-monoenergetic energy distribution under suitable injection conditions. The influence of transverse dynamics and space-charge effects is also examined, revealing additional constraints on acceleration efficiency associated with the transverse electromagnetic field of the driving microwave pulse. 
These results provide a quantitative assessment of the acceleration stage in microwave-driven plasma wakefield schemes and support their evaluation as a viable platform for compact plasma-based accelerators.
\end{abstract}
%============================================================================
\maketitle
%============================================================================
\section{Introduction}
%============================================================================
\noindent Plasma-based acceleration schemes have attracted sustained interest due to their ability to support accelerating electric fields far exceeding those achievable in conventional radiofrequency technologies~\cite{RevModPhys.81.1229,malka2012laser,esarey2002overview,wangler2008rf}. In these schemes, the collective plasma response generates longitudinal electric fields that trap and accelerate charged particles over short distances, enabling compact accelerator concepts~\cite{tajima1979laser,esarey2009physics}. While significant progress has been achieved in the generation of plasma wakefields driven by intense laser pulses or relativistic charged-particle beams, the practical performance of such schemes depends critically on a detailed understanding of particle injection, trapping conditions, and energy gain mechanisms~\cite{cakir2019brief,maity2025coupling,mirzaie2025progress,albert2016applications}.

\vspace{1pc}

\noindent In recent years, the excitation of plasma wakefields using high-power microwave pulses propagating in plasma-filled waveguides has emerged as a technologically attractive alternative to laser- or beam-driven approaches~\cite{aria2008wakefield,cao2024direct,lopez2025particle}. Advances in microwave sources have enabled the generation of subnanosecond pulses with peak powers in the gigawatt range and frequencies of gigahertz, parameters that are well suited for driving plasma waves in low-density plasmas~\cite{bliokh2017wakefield,krasik2019experiments,cao2019wakefield}. Compared with ultrashort laser systems, microwave-based schemes operate in a different parameter regime and may offer practical advantages in terms of technological accessibility and system complexity, albeit at acceleration gradients that are typically several orders of magnitude lower. However, despite growing experimental and theoretical efforts demonstrating the feasibility of microwave-driven wakefield generation, key questions remain regarding the efficiency of electron trapping and acceleration in such configurations.

\vspace{1pc}

\noindent In a previous work, fully electromagnetic three-dimensional particle-in-cell simulations were used to investigate the formation and properties of plasma wakefields excited by short ($\sim0.5$~ns), moderately intense ($\sim0.3$~GW) TE$_{10}$-mode microwave pulses propagating in rectangular plasma-filled waveguides~\cite{lopez2025particle}.  The wakefield response was characterized through systematic variations of key parameters, including pulse duration, frequency, power, waveguide geometry, and plasma density, revealing longitudinal electric field amplitudes on the order of kilovolts per centimeter. Building on those results, the present work focuses on the subsequent acceleration stage, examining how externally injected electrons interact with the sustained wakefield and identifying the conditions required for efficient trapping and energy gain.

\vspace{1pc}

\noindent The primary objective of this study is to analyze the dynamics of witness electrons injected into a microwave-driven plasma wakefield, with emphasis on identifying suitable initial conditions for capture, characterizing effective acceleration regions, and estimating the maximum attainable energy gain. To this end, the analysis is conducted in a set of complementary stages based on numerical simulations. First, a simplified numerical description is employed to simulate electron motion in a prescribed longitudinal wakefield, using wakefield parameters consistent with those reported in previous particle-in-cell simulations \cite{lopez2025particle}.This reduced framework provides physical insight into the trapping process and yields preliminary estimates of the expected energy gain. It also serves as a  practical guideline for selecting injection parameters and interpreting kinetic effects. In the second stage, electron acceleration is investigated using a reduced particle-in-cell framework in which the wakefield structure is obtained from fully electromagnetic simulations of the microwave–plasma interaction, while the injected witness electrons are treated as test particles. In this approach, the space-charge contribution of the witness bunch is neglected, so that the acceleration dynamics are governed solely by the electromagnetic fields of the wake and the driving microwave pulse.

\vspace{1pc}

\noindent Finally, fully self-consistent three-dimensional particle-in-cell simulations are performed to investigate electron acceleration under coupled electromagnetic and plasma dynamics. In this regime, the evolution of fields, background plasma, and injected electrons is resolved self-consistently, enabling a quantitative analysis of energy gain, energy spread, and the spatial evolution of the accelerated bunch.\\
By systematically combining reduced numerical descriptions with fully kinetic simulations, this work clarifies the role of injection conditions in microwave-driven plasma acceleration and supports the evaluation the potential of this scheme as a viable platform for compact plasma-based accelerators.
%============================================================================
\section{Physical Model and Numerical Approach}
\begin{figure}[ht]
    \centering
    \includegraphics[width=1.0\linewidth]{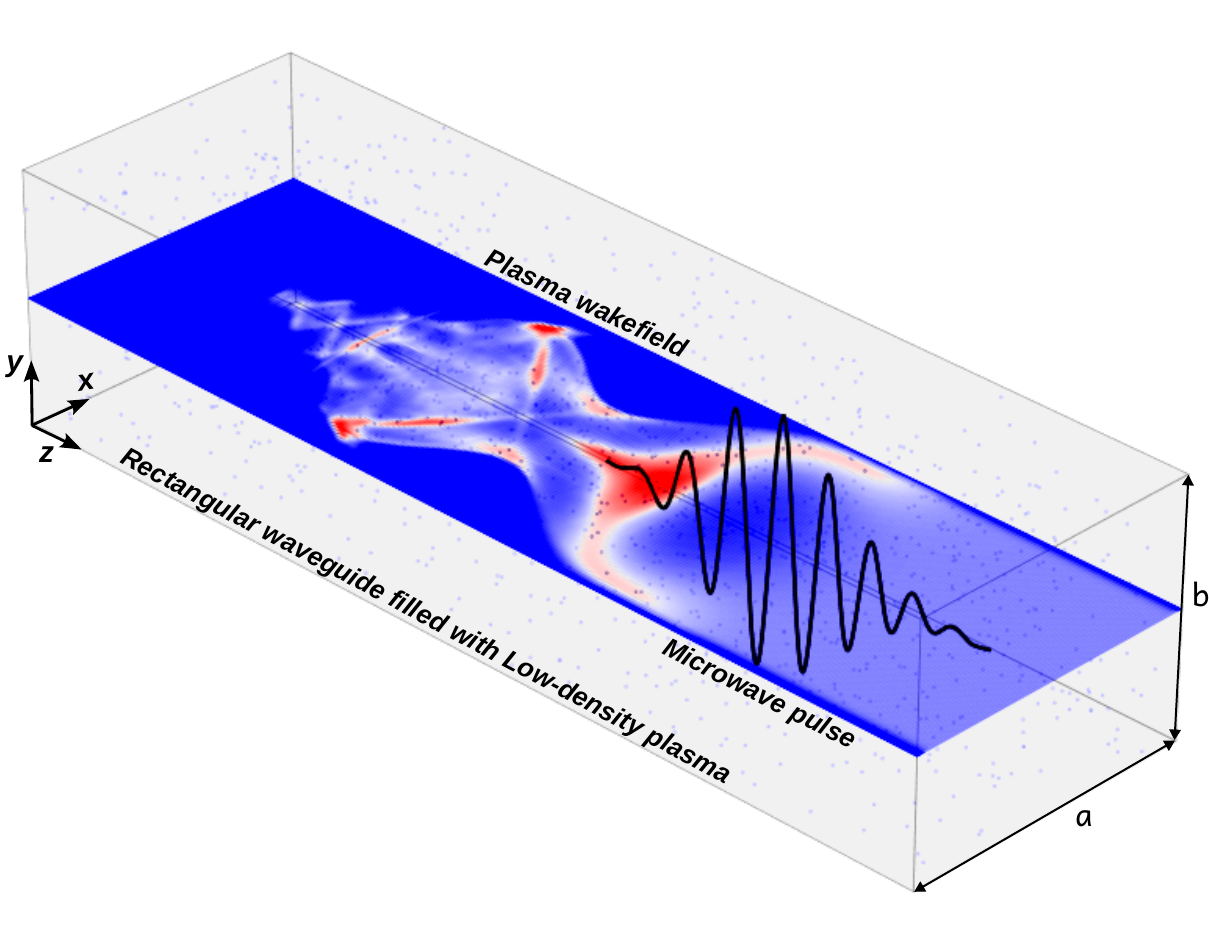}
    \caption{\label{fig_scheme}
    Schematic of the rectangular plasma-filled waveguide configuration. A short, high-power TE$_{10}$-mode microwave pulse propagates along the $z$ axis, exciting a longitudinal plasma wakefield trailing the driving pulse. The transverse waveguide dimensions are denoted by $a$ and $b$, and the Cartesian coordinate system $(x,y,z)$ is indicated.}
\end{figure}

%============================================================================
\noindent The system consists of a rectangular metallic waveguide with transverse dimensions $a = 3.0$~cm and $b = 2.1$~cm, filled with a cold plasma of initial electron density $n_0 = 1.8 \times 10^{10}$~cm$^{-3}$. The waveguide is excited by a short, high-power microwave pulse propagating along the longitudinal direction and polarized in the fundamental transverse electric mode TE$_{10}$. The driving pulse is characterized by a central frequency of 8~GHz, a temporal duration of approximately 0.44~ns, and a peak power of 0.25~GW. These parameters are selected to ensure propagation above the waveguide cutoff frequency and to enable efficient coupling of electromagnetic energy into the plasma, leading to the excitation of a longitudinal plasma wakefield. Figure~\ref{fig_scheme} presents a schematic representation of the physical configuration considered in this work. The rectangular waveguide geometry, plasma-filled region, and propagation direction of the TE$_{10}$ microwave pulse are illustrated, together with the adopted Cartesian coordinate system $(x,y,z)$, where the $z$-axis coincides with the direction of pulse propagation and wakefield excitation.

\noindent The present study is based on fully self-consistent three-dimensional particle-in-cell simulations~\cite{hockney1988computer,pukhov2015particle,lapenta2015kinetic}. The electromagnetic solver employs the current deposition scheme proposed by Umeda \textit{et al.}, which ensures exact charge conservation throughout the simulation domain~\cite{umeda2003new}. Within this formulation, the coupled evolution of the electromagnetic fields, the background plasma response, and the relativistic dynamics of the injected electrons is resolved self-consistently~\cite{birdsall2004plasma,pukhov2015particle}. This approach enables a quantitative analysis of electron acceleration, including energy gain, phase-space evolution, and spatial dynamics within the plasma wakefield.

\vspace{1pc}

\noindent The metallic waveguide walls located at $x = 0$, $x = a$, $y = 0$, and $y = b$ are modeled using perfect electric conductor (PEC) boundary conditions. To reduce numerical noise and spurious reflections at the microwave injection boundary, the plasma density is smoothly increased from vacuum to its nominal value over a short entrance region along the $z$-axis, providing a gradual transition into the homogeneous plasma bulk. The TE$_{10}$-mode microwave pulse is initialized at $z = 0$, and its temporal propagation and interaction with the plasma are resolved self-consistently.

\vspace{1pc}

\noindent Macroparticles are subject to absorbing boundary conditions at the metallic waveguide walls. A two-cell absorbing layer is implemented adjacent to each transverse boundary to suppress unphysical particle reflections. When a macroparticle enters this layer, its trajectory is advanced up to the intersection with the boundary within the current time step, and its charge and current contributions are deposited consistently using the same charge-conserving scheme employed in the bulk plasma. The particle is then removed from the simulation domain.

\vspace{1pc}

\noindent The spatial grid is defined by $\Delta x = \Delta y = a/100$, where $a$ denotes the waveguide width, and $\Delta z = \lambda_0/32$, providing approximately 33 grid points per wavelength $\lambda_0$ of the injected microwave pulse. This spatial resolution ensures provides adequate sampling of both the electromagnetic wavelength of the driving pulse and the plasma wavelength associated with the excited wakefield. The temporal step is chosen to satisfy the Courant–Friedrichs–Lewy (CFL) stability condition for electromagnetic particle-in-cell simulations~\cite{sullivan2013electromagnetic,courant1967partial}.

\vspace{1pc}

\noindent Overall, this computational framework resolves the key nonlinear and kinetic processes governing microwave-driven plasma wakefields and the subsequent acceleration of externally injected electrons. It therefore enables a systematic evaluation of this scheme as a potential building block for compact plasma-based accelerators.
%============================================================================
\section{Results and Discussion}
%============================================================================
\noindent This section presents and discusses the results of the electron acceleration study, which is organized in two stages. First, a reduced numerical description is used to identify suitable initial conditions for electron injection and to obtain preliminary estimates of the expected energy gain in the microwave-driven wakefield. These results provide physical insight into the trapping and acceleration mechanisms and serve as a reference for the second stage. Subsequently, self-consistent three-dimensional particle-in-cell simulations are employed to investigate electron acceleration under realistic electromagnetic and plasma conditions, enabling for a quantitative assessment of energy gain and beam dynamics.
%---------------------------------------------------------------------------
\subsection{Reduced numerical analysis of electron injection}
%---------------------------------------------------------------------------
\noindent The first stage of the analysis focuses on determining suitable initial conditions for electron injection into the microwave-driven plasma wakefield and on estimating the corresponding energy gain. In this stage, the wakefield structure obtained from previous fully electromagnetic three-dimensional particle-in-cell simulations is treated as a prescribed accelerating field, and the dynamics of injected electrons are analyzed using a reduced numerical description. These simulations showed that the interaction between the microwave pulse and the plasma leads to the formation of a coherent plasma wakefield trailing the driving pulse. The wakefield is characterized by a longitudinal electric field with peak amplitudes on the order of $E_z \sim 1$~kV/cm, followed by damped oscillations. In this configuration, the negative phase of the wake corresponds to an accelerating field for electrons propagating in the direction of the microwave pulse. In the present work, this reference wakefield configuration serves as the accelerating structure for analyzing electron injection, trapping, and energy gain in a controlled manner, isolating the role of the longitudinal electric field prior to addressing collective and self-consistent effects.

\vspace{1pc}

\noindent As a starting point, a simplified one-dimensional numerical description is adopted, motivated by the fact that electron acceleration predominantly occurs along the central axis of the waveguide. In this framework, the plasma wakefield is represented by a damped sinusoidal longitudinal electric field with a peak amplitude of $E_z = 1.2$~kV/cm and a spatial period equal to the plasma wavelength $\lambda_p$, corresponding to a reference plasma density of $n_0 = 1.8 \times 10^{10}$~cm$^{-3}$. The wakefield is assumed to propagate with a group velocity $v_g = 0.77\,c$, consistent with the propagation speed of the 8~GHz microwave pulse in a rectangular waveguide with dimensions $a = 3$~cm and $b = 0.7\,a$, in agreement with previously reported particle-in-cell simulations~\cite{lopez2025particle}.

\vspace{1pc}

\begin{figure}[ht]
    \centering
    \includegraphics[width=0.6\linewidth]{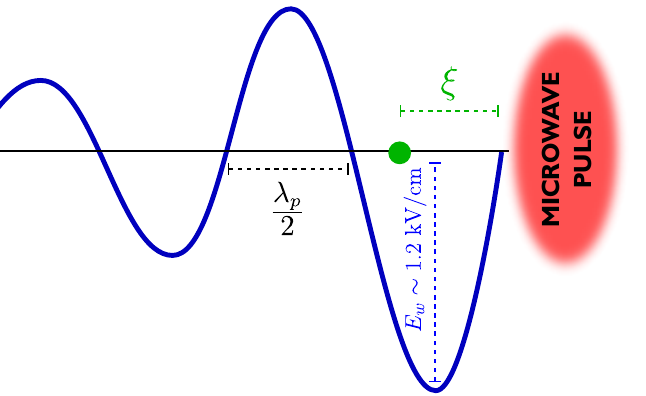}
    \caption{\label{fig_injection_scheme}
    One-dimensional schematic showing the longitudinal electric field $E_z$ of the microwave-driven plasma wakefield. The spatial coordinate $\xi$ denotes the position relative to the wake phase. A witness electron (green marker) is injected within the first accelerating region, corresponding to the negative phase ($E_z < 0$), where efficient trapping and  energy gain occur.}
\end{figure}

\noindent Figure~\ref{fig_injection_scheme} illustrates the injection scheme adopted in the reduced description, in which a test electron is injected at a longitudinal position $\xi$ within the first accelerating bucket of the wakefield. Regions of accelerating ($E_z < 0$) and decelerating ($E_z > 0$) longitudinal electric field are indicated, enabling the identification of favorable injection phases for electron trapping and energy gain.
\vspace{1pc}

\noindent Test electrons are injected at different longitudinal positions $\xi$ within the first accelerating bucket of the wakefield. For each injection position, a range of initial longitudinal velocities $v_{z_0}$ is explored. The subsequent evolution of the electron motion under the prescribed wakefield is analyzed, and the maximum longitudinal kinetic energy gain $\Delta K$ attained during the interaction is evaluated.

\vspace{1pc}

\begin{figure}[ht]
    \centering
    \includegraphics[width=0.7\linewidth]{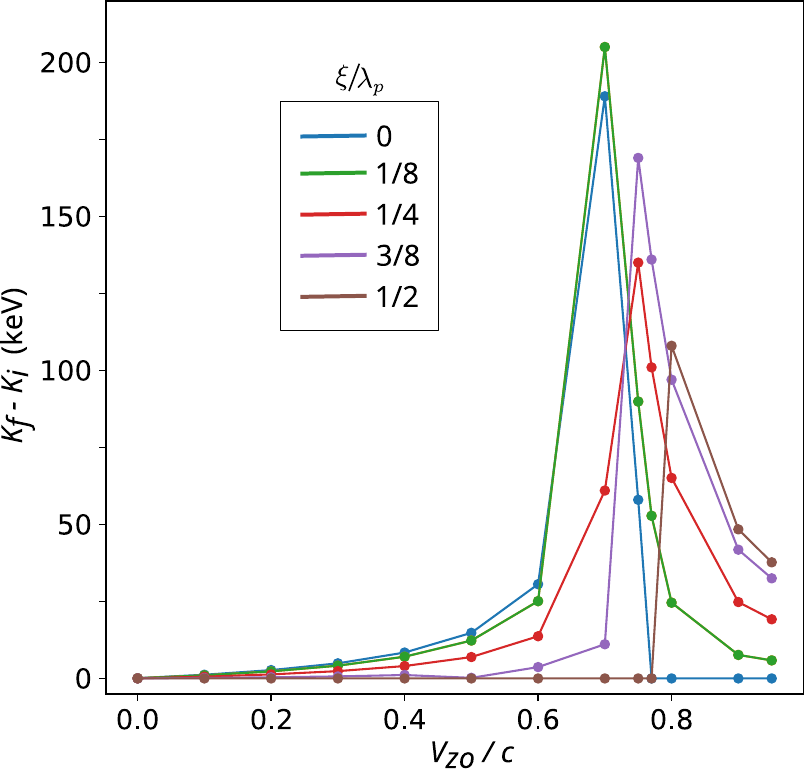}
    \caption{\label{fig_injection_results}
    Energy gain of the witness electron as a function of the initial longitudinal velocity $v_{z0}$ for five representative injection positions $\xi$ within the first accelerating bucket of the microwave-driven plasma wakefield. For each injection phase, an optimal initial velocity is observed.}
\end{figure}

\noindent Figure~\ref{fig_injection_results} shows the resulting energy gain as a function of the initial electron velocity for five representative injection positions within the first wakefield bucket. For each value of $\xi$, an optimal initial velocity is observed, corresponding to the most efficient phase-locking between the electron and the propagating wake. The acceleration efficiency depends strongly on the injection phase. The most favorable case is obtained for $\xi = \lambda_p/8$ and $v_{z_0} = 0.7\,c$, yielding a maximum energy gain of approximately 200~keV over an interaction length of about 2~m.

\noindent A key feature of these results is that the optimal initial velocities are clustered around the group velocity of the driving microwave pulse ($v_g \sim 0.77\,c$). This implies that the witness electrons must be pre-accelerated prior to injection into the wakefield. In the optimal case identified here, an initial kinetic energy of approximately 205~keV is required, allowing the electron to reach a final energy close to 400~keV. Table~\ref{Tab_acc1} summarizes the optimal initial velocities, maximum energy gain, ratio between final and initial kinetic energies $K_f/K_0$, and effective interaction length $d_z$ for the different injection positions considered.

\begin{table}[ht]
\begin{center}
\begin{tabular}{ccccc}
\hline \hline
\(\xi/\lambda_p\) & \(v_{z_0}/c\) & \(\Delta K\) (keV) & \(K_f/K_0\) & \(d_z\) (m) \\
\hline
0   & 0.70 & 189 & 1.92 & 2.06 \\
1/8 & 0.70 & 205 & 2.00 & 2.04 \\
1/4 & 0.75 & 135 & 1.52 & 1.41 \\
3/8 & 0.75 & 169 & 1.65 & 2.03 \\
1/2 & 0.80 & 108 & 1.35 & 1.53 \\
\hline \hline
\end{tabular}
\caption{\label{Tab_acc1}
Optimal initial longitudinal velocities $v_{z_0}$, maximum kinetic energy gain $\Delta K$, ratio of final to initial kinetic energy $K_f/K_0$, and effective interaction length $d_z$ for five injection positions $\xi$ within the first accelerating bucket of the microwave-driven plasma wakefield. The most efficient coupling occurs for $\xi =\lambda_p/8$, leading to a final electron energy close to 400~keV.}
\end{center}
\end{table}

\noindent This reduced numerical analysis provides a quantitative estimate of the achievable energy gain and identifies favorable injection conditions under the action of the longitudinal wakefield alone. These results serve as a reference for initializing the witness electrons in the fully self-consistent particle-in-cell simulations discussed in the following subsection.
%---------------------------------------------------------------------------
\subsection{Electron dynamics in the test-particle regime}
%---------------------------------------------------------------------------
\noindent Once favorable injection conditions for external electrons in the microwave-driven plasma wakefield have been identified, the dynamics of an injected electron bunch is analyzed under the test-particle approximation. In this regime, the bunch evolves under the combined action of the electromagnetic fields of the wakefield and the driving microwave pulse, while space-charge effects are neglected. Consequently, the injected electrons do not modify the wakefield structure and generate significant collective fields. This approach isolates the role of transverse electromagnetic effects and extends the previous analysis, which was limited to the longitudinal wakefield component.

\vspace{1pc}

\begin{figure}[ht]
    \centering
    \includegraphics[width=1.0\linewidth]{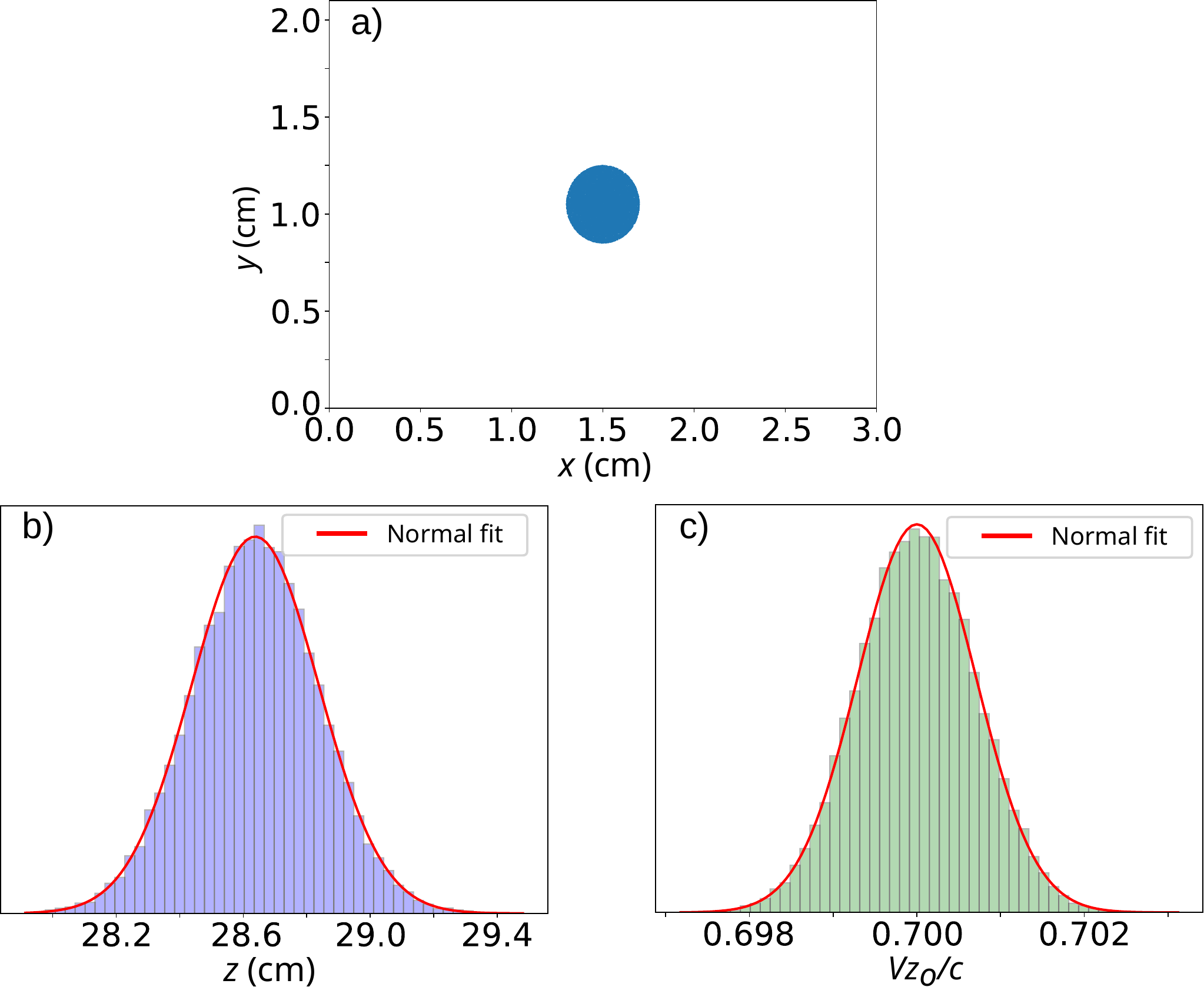}
    \caption{\label{fig_bunch_inicial}
    Initial phase-space distributions of the injected electron bunch. 
    (a) Transverse distribution in the $x$–$y$ plane, showing a circular cross section with radius $r_b = 2$~mm. 
    (b) Longitudinal distribution characterized by $\sigma_z = 2$~mm. 
    (c) Longitudinal velocity distribution with a relative spread of $0.1\%$. 
    Gaussian profiles are adopted in all cases.}
\end{figure}

\noindent In this stage, a witness bunch is injected along the central axis of the waveguide, where the wakefield amplitude is maximal, and at the optimal longitudinal injection phase $\xi = \lambda_p/8$. The bunch has a circular transverse cross section with radius $r_b = 2$~mm and a longitudinal extent characterized by $\sigma_z = 2$~mm. The initial longitudinal velocity is set to $v_z = 0.7\,c$, with a relative velocity spread of $0.1\%$, corresponding to an initial energy spread of approximately $0.34\%$. The initial transverse velocities are set to zero. The complete set of injection parameters is summarized in Table~\ref{tab_param_bunch_PS}, and the corresponding initial distributions are shown in Fig.~\ref{fig_bunch_inicial}, exhibiting Gaussian profiles.

\begin{table}[ht]
\begin{ruledtabular}
\begin{tabular}{lcc}
\textbf{Parameter} & \textbf{Symbol} & \textbf{Value} \\
\hline
Injection position & $\xi$ & $\lambda_p/8$ \\
Transverse radius & $r_b$ & $2$~mm \\
Longitudinal extent & $\sigma_z$ & $2$~mm \\
Initial velocity & $v_z$ & $0.7\,c$ \\
Relative velocity spread & $\Delta v_z / v_z$ & $0.1\%$ \\
Initial energy spread & $\Delta K / K$ & $0.34\%$ \\
Initial transverse velocities & $v_x,\,v_y$ & $0$ \\
\end{tabular}
\caption{\label{tab_param_bunch_PS}
Initial parameters of the injected electron bunch employed in the test-particle simulations.}
\end{ruledtabular}
\end{table}

\begin{figure}[ht]
    \centering
    \includegraphics[width=0.9\linewidth]{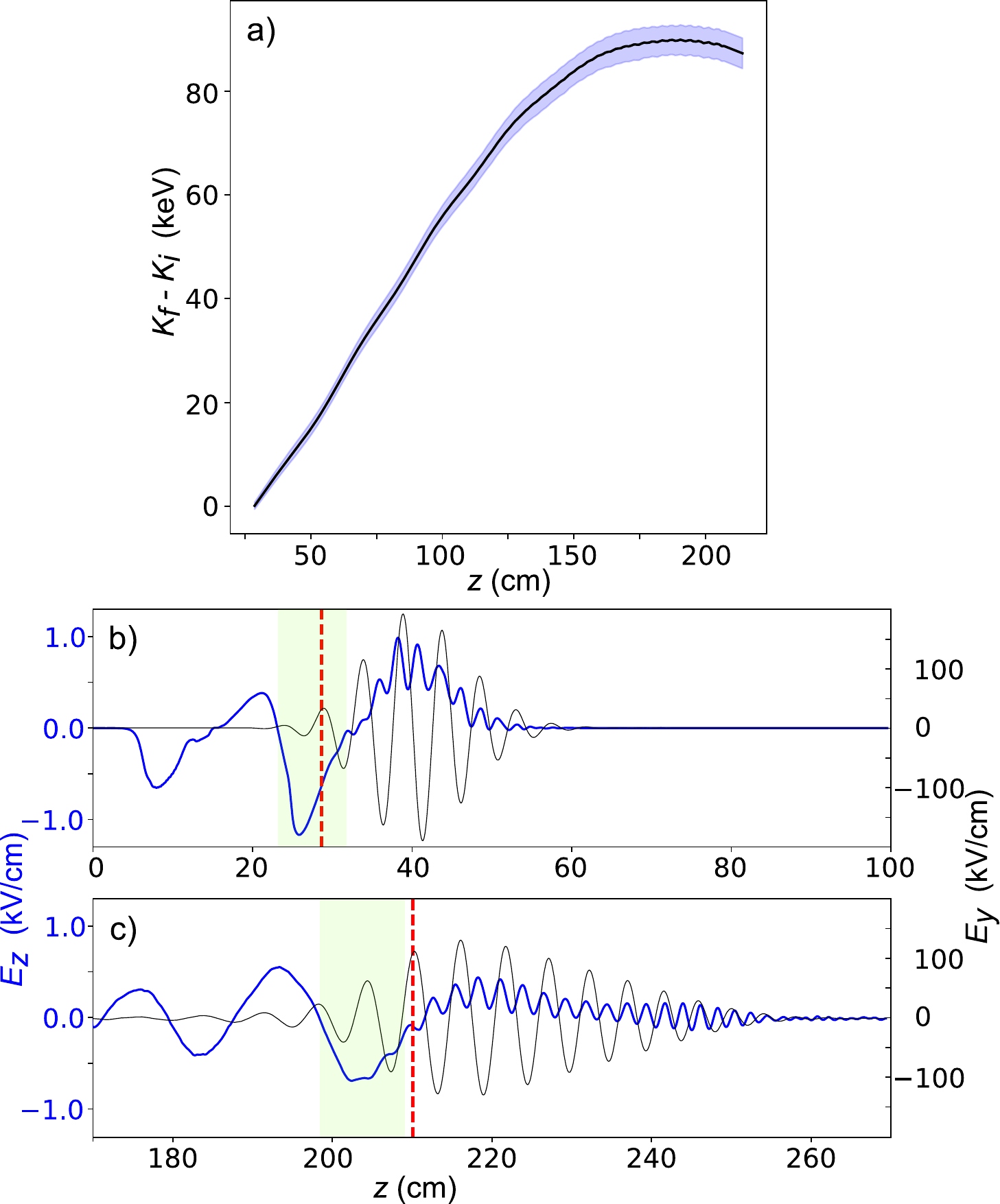}
    \caption{\label{fig_acc_sp}
    (a) Mean energy gain (black line) and associated energy spread $\Delta K$ (blue shaded area) of the electron bunch in the test-particle regime. 
    (b) Longitudinal wakefield $E_z$ (blue) and transverse microwave field $E_y$ (black) at the injection stage. 
    (c) Same field components as in (b) at the end of the simulation. 
    The red dashed line indicates the mean longitudinal position of the bunch, and the green shaded region denotes the effective accelerating phase of the wakefield.}
\end{figure}

\noindent The injected bunch is accelerated by the wakefield; however, the maximum energy gain is lower than that predicted by the one-dimensional analysis. Figure~\ref{fig_acc_sp}(a) shows the evolution of the mean energy gain and the corresponding energy spread. The average energy gain reaches a maximum value of approximately 90~keV, while the relative energy spread increases to $\Delta K/K \sim 0.9\%$, indicating that the bunch remains quasi-monoenergetic at the end of the interaction.

\vspace{1pc}

\noindent The primary limitation on the achievable energy gain arises from the longitudinal slippage of the electrons into regions of the wakefield where the electric field changes sign, leading to partial deceleration. This behavior is illustrated in Fig.~\ref{fig_acc_sp}(c), where the mean longitudinal position of the bunch (red dashed line) exits the effective accelerating region (green shaded area). In addition, the wakefield amplitude decreases as the microwave pulse disperses during propagation, as a consequence of its finite spectral bandwidth $\Delta f$, as shown in Fig.~\ref{fig_acc_sp}.

\begin{figure}[ht]
    \centering
    \includegraphics[width=0.70\linewidth]{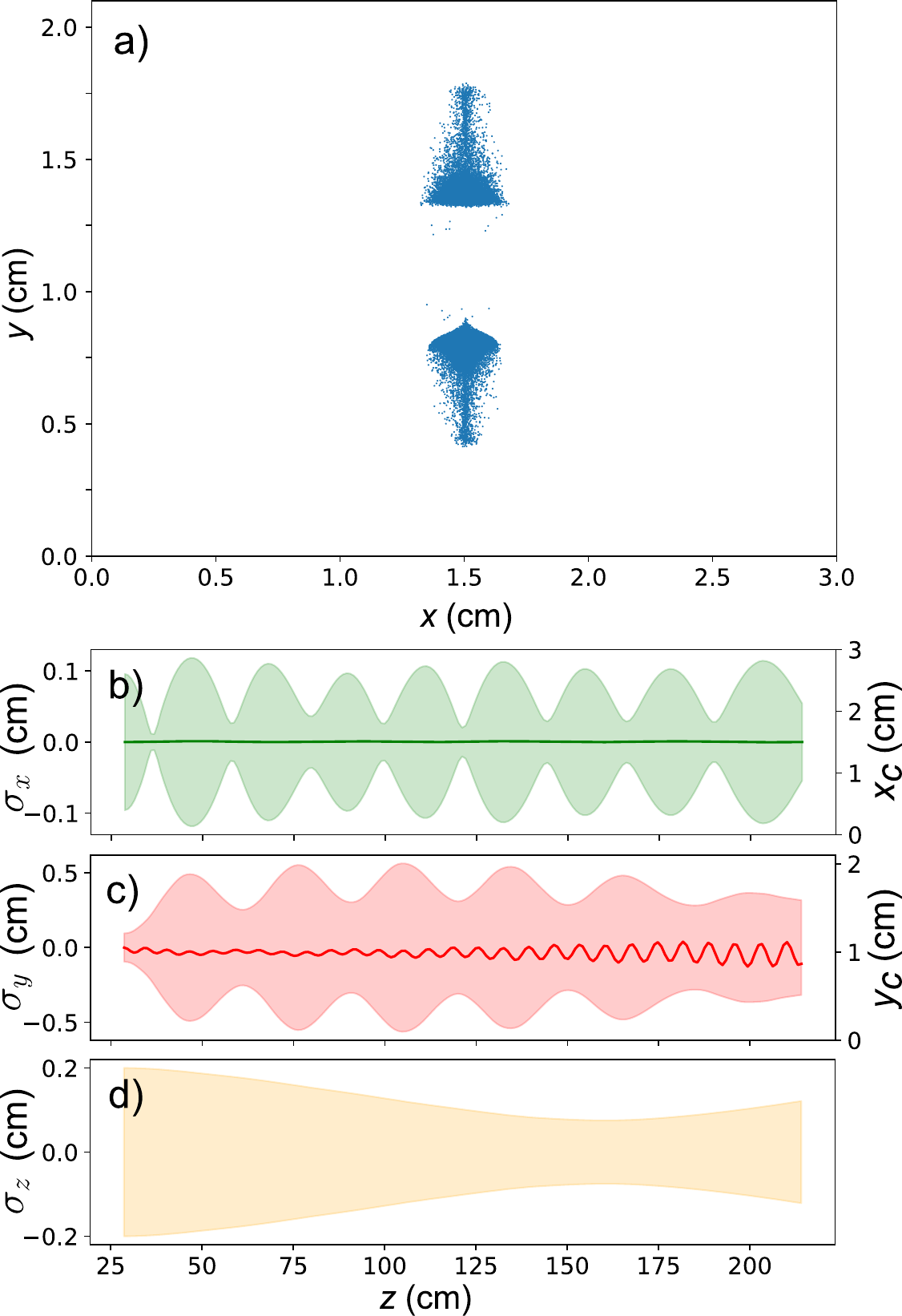}
    \caption{\label{fig_acc_sp_xy}
    (a) Transverse distribution of the electron bunch at an intermediate stage of acceleration ($z \sim 118$~cm). 
    (b)–(d) Evolution of the mean transverse positions $(x_c, y_c)$ and the corresponding standard deviations $(\sigma_x, \sigma_y, \sigma_z)$ along the propagation direction. 
    The shaded regions represent the spatial dispersion of the bunch during its interaction with the wakefield and the driven microwave pulse.}
\end{figure}

\noindent In addition to the reduction in wakefield amplitude, a second mechanism affects the acceleration efficiency: transverse distortion of the electron bunch during acceleration. In particular, the bunch exhibits pronounced deformation along the $y$ axis, as shown in Fig.~\ref{fig_acc_sp_xy}. Figure~\ref{fig_acc_sp_xy}(a) displays the transverse particle distribution at an intermediate stage of the acceleration, corresponding to a central longitudinal position $z_c \sim 118$~cm. Figures~\ref{fig_acc_sp_xy}(b)–\ref{fig_acc_sp_xy}(d) show the evolution of the mean transverse positions $(x_c, y_c)$ and the associated standard deviations $(\sigma_x, \sigma_y, \sigma_z)$, which characterize the spatial dispersion of the bunch.

\noindent Along the $x$ axis, the bunch exhibits only moderate oscillations and remains confined near the waveguide axis, with a transverse size close to its initial value, $\sigma_x = 2$~mm. In contrast, the pronounced distortion along the $y$-direction can be directly attributed to the transverse electric field component of the TE$_{10}$ mode, which is polarized exclusively along $y$. This interpretation is supported by Fig.~\ref{fig_acc_sp}(b), where the microwave field amplitude at the bunch location exceeds that of the wakefield by approximately one order of magnitude during the early stages of acceleration. As the bunch approaches the dispersing microwave pulse, the transverse interaction is further enhanced, leading to oscillatory motion of the bunch centroid around the waveguide midplane.

\vspace{1pc}

\noindent In this context, the ponderomotive force associated with the microwave pulse must also be taken into account. On average, this force acts opposite to the longitudinal motion of the electrons, introducing an effective decelerating contribution that competes with the accelerating action of the wakefield. The combined effects of wakefield acceleration and ponderomotive slowing therefore constitutes an additional limitation on the maximum attainable energy gain.

\vspace{1pc}

\noindent Finally, Fig.~\ref{fig_acc_sp_xy}(d) shows that the longitudinal bunch length $\sigma_z$ decreases progressively up to the point of maximum energy gain, indicating longitudinal compression of the electron bunch. Beyond this point, the bunch begins to expand again as the interaction with the wakefield loses coherence. This behavior results from the interplay between the accelerating wakefield and the opposing ponderomotive force of the microwave pulse.

\vspace{1pc}

\noindent Overall, this stage of the analysis identifies the dominant physical mechanisms governing bunch dynamics in the test-particle regime: (i) acceleration driven by the longitudinal wakefield and (ii) transverse and longitudinal limitations introduced by the dispersing microwave pulse and its associated ponderomotive effects. These results provide a reference for the fully self-consistent particle-in-cell simulations presented in the following subsection, where space-charge effects and field back-reaction are included.
%---------------------------------------------------------------------------
\subsection{Self-consistent dynamics of electron acceleration}
%---------------------------------------------------------------------------
\noindent This subsection presents the results corresponding to the fully self-consistent dynamics of the electron acceleration process, in which all relevant interaction mechanisms are simultaneously taken into account. In contrast to the test-particle approximation discussed previously, the collective effects associated with the injected electron bunch are included here. As a result, the bunch not only experiences the electromagnetic fields of the wakefield and the driving microwave pulse, but also modifies the excited fields through its own space-charge contribution. This approach provides a more realistic description of the acceleration process and allows the net impact of collective effects on acceleration efficiency and beam quality to be assessed.

\vspace{1pc}

\noindent The acceleration of an electron bunch with a total charge of 50~pC is simulated, while maintaining a longitudinal extent of $\sigma_z \approx 2\mathrm{mm}$ and a transverse radius of 0.3~mm. The initial longitudinal velocity is set to $v_{z_0} = 0.7\,c$, with a relative velocity spread of $0.1\%$, and the initial transverse velocities are set to zero. The bunch is injected along the central axis of the waveguide at a time such that the wakefield phase allows injection at the optimal relative position $\xi = \lambda_p/8$.

\vspace{1pc}

\begin{figure}[ht]
    \centering
    \includegraphics[width=1.0\linewidth]{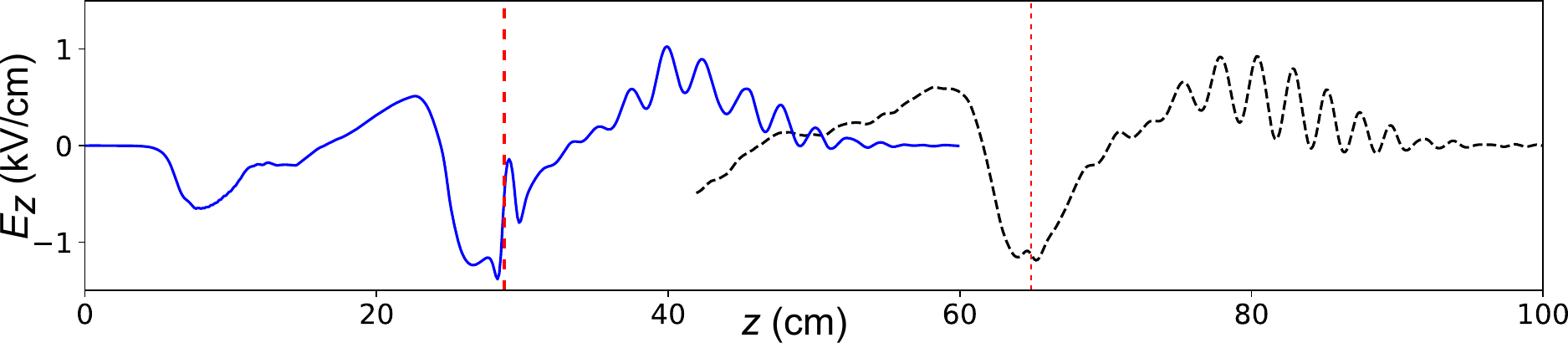}
    \caption{\label{fig_wake_50pc_sc}
    Longitudinal electric field $E_z$ illustrating the deformation of the plasma wakefield during the injection of a 50~pC electron bunch. The blue curve corresponds to the field immediately after injection, showing a localized perturbation at the bunch position. The black dashed curve represents the wakefield at a later time, indicating partial recovery of the wake structure. Vertical lines mark the longitudinal position of the bunch in each case.}
\end{figure}

\noindent Under these conditions, the injection of the electron bunch does not lead to a significant disruption of the wakefield structure. This behavior is illustrated in Fig.~\ref{fig_wake_50pc_sc}, where a localized perturbation of the longitudinal electric field $E_z$ is observed at the injection stage (blue curve), coincident with the bunch position. As the system evolves, this perturbation is progressively attenuated, and the wakefield structure is largely recovered at later times (black dashed curve), indicating that collective effects remain moderate for the considered bunch charge.

\begin{figure}[ht]
    \centering
    \includegraphics[width=0.70\linewidth]{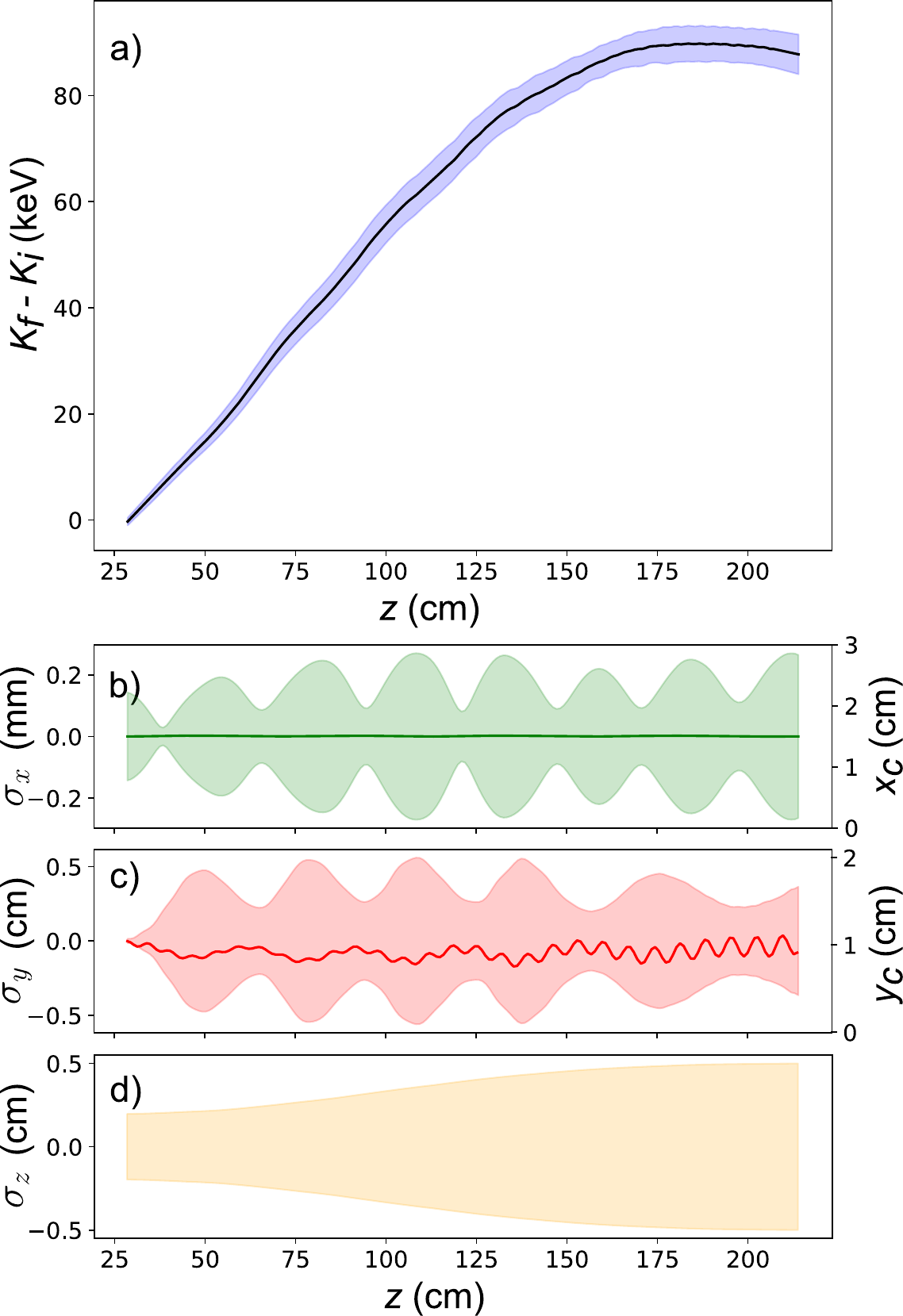}
    \caption{\label{fig_acc_50pc_sc}
    (a) Evolution of the average energy gain of the accelerated electron bunch; the shaded region indicates the corresponding energy spread. 
    (b)–(d) Evolution of the bunch size (shaded regions) and centroid positions $x_c$ and $y_c$ (solid lines) along the $x$, $y$, and $z$ directions during propagation in the waveguide.}
\end{figure}

\noindent The results shown in Fig.~\ref{fig_acc_50pc_sc}(a) reveal an average energy gain of approximately 90~keV, in good agreement with the estimate obtained from the test-particle analysis. At the point of maximum energy gain, the relative energy spread remains below 1\%, preserving the quasi-monoenergetic character of the accelerated bunch. Figures~\ref{fig_acc_50pc_sc}(b)–\ref{fig_acc_50pc_sc}(d) show the evolution of the spatial extent of the bunch along the transverse and longitudinal directions, where the shaded regions represent the standard deviations $\sigma_x$, $\sigma_y$, and $\sigma_z$. Along the $x$-direction, the bunch width increases by approximately a factor of two, but remains on the order of a few tenths of a millimeter, indicating overall transverse confinement. In contrast, a significantly stronger deformation is observed along the $y$-direction, where the effective bunch width increases from sub-millimeter values at injection to approximately 1~cm at later stages of the interaction. This pronounced transverse spreading is attributed to the interaction with the tail of the microwave pulse, whose transverse electric field component $E_y$ is substantially stronger than the wakefield amplitude and therefore dominates the transverse dynamics.

\noindent The evolution of the transverse centroid positions further supports this interpretation. While the mean position $x_c$ remains closely centered around the waveguide axis throughout the interaction, the $y$-centroid exhibits more pronounced oscillations, consistent with the strong influence of the microwave field polarized along the $y$-direction. These oscillations lead to a slight deviation from the waveguide midplane, although no macroscopic beam loss is observed.

\noindent Finally, in contrast to the test-particle regime discussed in the previous subsection, no longitudinal compression of the bunch is observed in the self-consistent simulations. Instead, the bunch length increases progressively, reaching approximately twice its initial value. This behavior is attributed to internal repulsive forces associated with space-charge effects, which counteract any longitudinal focusing induced by the wakefield. These results highlight the role of collective effects in limiting beam quality and underscore the importance of self-consistent modeling for realistic assessments of microwave-driven plasma acceleration.
%---------------------------------------------------------------------------
\subsection{Influence of injection phase on electron acceleration}
%---------------------------------------------------------------------------
\noindent The influence of the injection phase on the acceleration dynamics of the electron bunch is analyzed by varying the initial injection position while keeping all other bunch parameters identical to those of the reference self-consistent case. In particular, the bunch charge, longitudinal duration, transverse size, initial velocity, and energy spread are held fixed, and only the relative injection phase $\xi$ within the wakefield is modified. Two additional injection scenarios are considered: injection at the location of maximum accelerating field, $\xi = \lambda_p/4$, and injection at an intermediate phase, $\xi = 3\lambda_p/8$. According to the reduced one-dimensional analysis discussed previously, both cases are expected to yield lower energy gains. Here, their impact is examined within a fully self-consistent framework to assess the resulting acceleration efficiency and beam quality.

\vspace{1pc}

\begin{figure}[ht]
    \centering
    \includegraphics[width=1.0\linewidth]{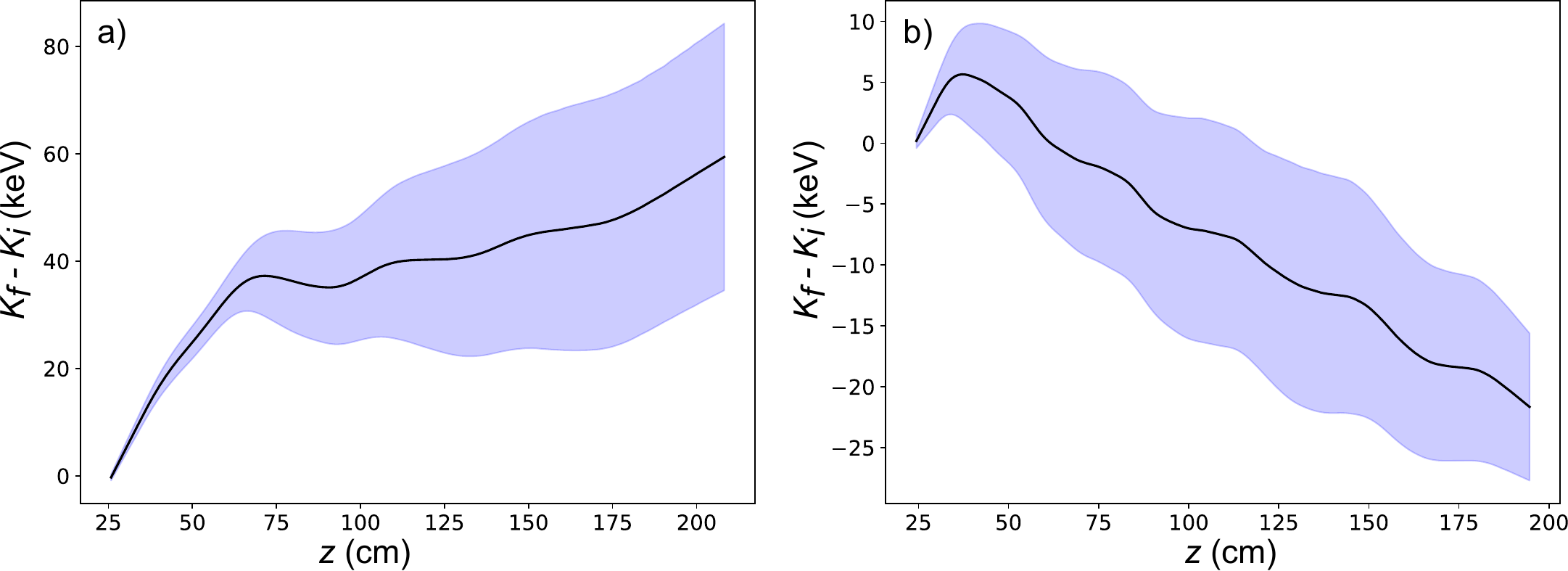}
    \caption{\label{fig_energias_vs_xi}
    Evolution of the mean kinetic energy of the electron bunch for two injection phases: (a) $\xi = \lambda_p/4$ and (b) $\xi = 3\lambda_p/8$. Shaded regions indicate the corresponding energy spread in each case.}
\end{figure}

\noindent The simulation results confirm these expectations. For injection at $\xi = \lambda_p/4$, the electron bunch experiences a reduced energy gain of approximately 60~keV, as shown in Fig.~\ref{fig_energias_vs_xi}(a). The case $\xi = 3\lambda_p/8$ proves even less favorable, as illustrated in Fig.~\ref{fig_energias_vs_xi}(b): after a short acceleration stage, with a maximum energy gain of only about 5~keV, the bunch enters a decelerating phase and ultimately loses more than 20~keV relative to its initial energy.

\noindent In addition to the reduced net energy gain, injection at $\xi = \lambda_p/4$ leads to a significantly larger energy spread compared to the optimal case discussed in the previous subsection. As a result, the accelerated bunch no longer exhibits a quasi-monoenergetic character, indicating a degradation of beam quality even in cases where a net energy increase is achieved.

\vspace{1pc}

\begin{figure}[ht]
    \centering
    \includegraphics[width=1.0\linewidth]{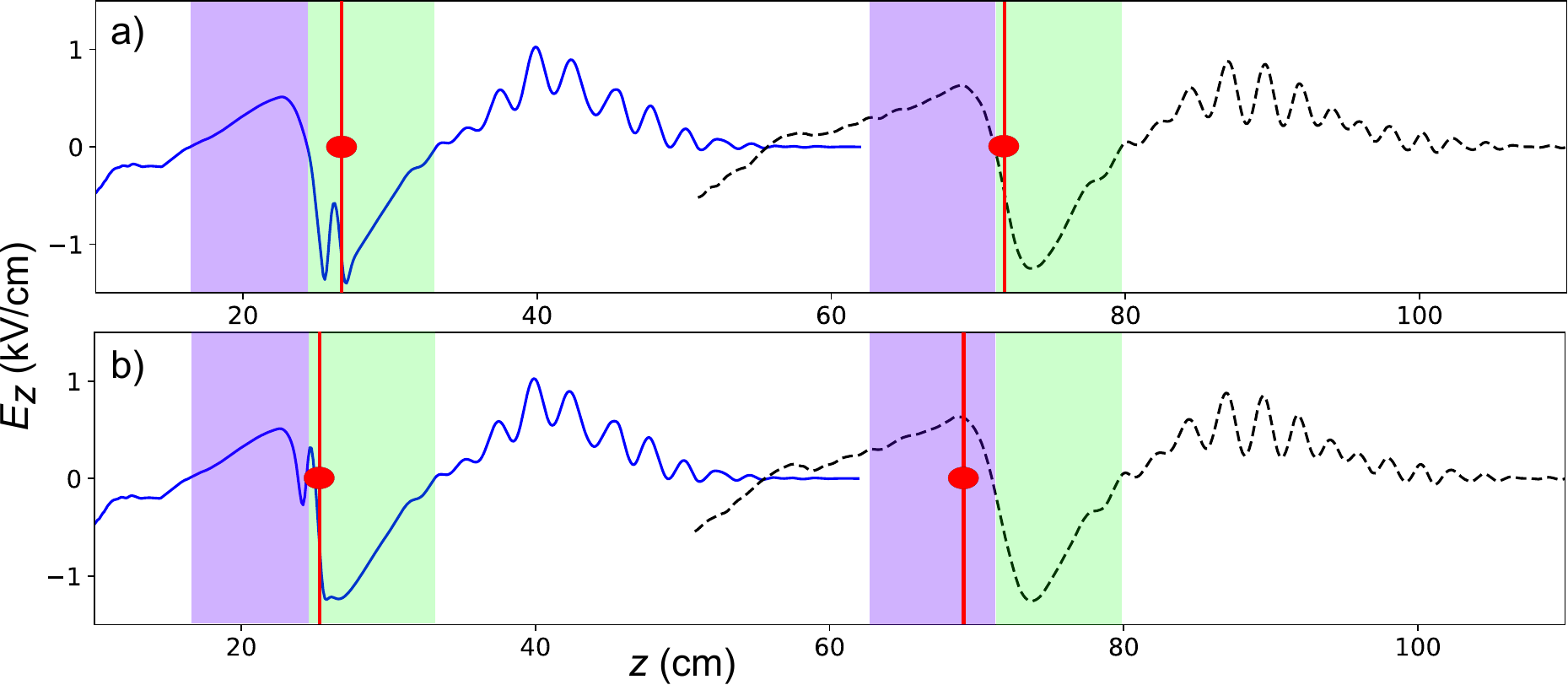}
    \caption{\label{FIG_wake_vs_xi}
    Evolution of the longitudinal position of the electron bunch relative to the wakefield structure for two injection phases: (a) $\xi = \lambda_p/4$ and (b) $\xi = 3\lambda_p/8$. The blue curve shows the longitudinal electric field $E_z$ immediately after injection, while the black dashed curve corresponds to the field after approximately 45~cm of bunch propagation. The vertical line and red elliptical region indicate the centroid position and approximate longitudinal extent of the bunch. Green- and purple-shaded regions denote accelerating ($E_z < 0$) and decelerating ($E_z > 0$) phases, respectively.}
\end{figure}

\noindent The underlying physical mechanism responsible for this behavior becomes evident when examining the relative position of the bunch with respect to the wakefield structure. In the least favorable case, $\xi = 3\lambda_p/8$, the bunch is injected near the trailing edge of the accelerating region, that is, close to the boundary between acceleration and deceleration. Since the initial injection velocity ($v_{z0} = 0.7\,c$) is lower than the wake phase velocity ($\sim 0.77\,c$), the bunch rapidly slips into the decelerating phase, as illustrated in Fig.~\ref{FIG_wake_vs_xi}(b), thereby preventing sustained energy gain.

\vspace{1pc}

\noindent For injection at $\xi = \lambda_p/4$, the bunch initially occupies the center of the accelerating bucket. However, as the wake propagates, longitudinal phase slippage combined with finite energy and velocity spreads causes a fraction of electrons to cross into the decelerating region. This leads to a reduction in net energy gain and a substantial increase in energy spread. The resulting dynamics are illustrated in Fig.~\ref{FIG_wake_vs_xi}, which compares the evolution of the bunch centroid relative to the wakefield structure for the two injection phases considered.

\noindent Overall, these results highlight the critical role of the injection phase in determining both the achievable energy gain and the final beam quality in microwave-driven plasma wakefield acceleration. Even within a fully self-consistent framework, deviations from the optimal injection phase lead to substantial reductions in acceleration efficiency and, in some cases, to the complete suppression of net energy gain. These findings reinforce the conclusions drawn from the reduced numerical analysis and underscore the necessity of precise phase control for effective electron acceleration in this scheme.
%============================================================================
\section{Summary and Conclusions}
%============================================================================
\noindent In this work, the acceleration dynamics of externally supplied electrons in microwave-driven plasma wakefields sustained in rectangular plasma-filled waveguides has been investigated using three-dimensional particle-in-cell simulations. Building upon previously reported results on wakefield formation, the present study focuses on the acceleration stage, analyzing the conditions under which electrons already embedded in the accelerating phase of the wake can gain energy, as well as the influence of transverse dynamics and collective effects.

\medskip

\noindent A reduced numerical analysis was first employed to identify favorable injection parameters. This simplified description revealed a strong dependence of the achievable energy gain on the injection phase and initial velocity. Optimal acceleration occurs when electrons are injected within the first accelerating bucket with an initial velocity close to the wake phase velocity, enabling sustained phase synchronization with the propagating structure. Under these conditions, energy gains on the order of $100$~keV over meter-scale interaction lengths were predicted.

\medskip

\noindent Test-particle simulations of a finite-size electron bunch demonstrated that transverse electromagnetic fields associated with the microwave pulse significantly influence beam dynamics. In particular, the transverse electric component of the TE$_{10}$ mode induces anisotropic deformation of the bunch, especially along the polarization direction, reducing the effective acceleration efficiency compared with the one-dimensional estimate.

\medskip

\noindent Fully self-consistent simulations including space-charge effects confirmed that, for moderate bunch charges, the global structure of the wakefield remains largely preserved, and net energy gains comparable to the test-particle regime are obtained. However, collective effects lead to progressive longitudinal broadening and enhanced transverse distortion of the electron bunch, thereby degrading beam quality even when significant energy gain is achieved.

\medskip

\noindent The sensitivity of the acceleration process to the injection phase was further examined by considering off-optimal injection positions. These simulations showed that small deviations from the optimal phase substantially reduce energy gain and may even lead to net deceleration, highlighting the importance of precise phase control in microwave-driven plasma acceleration schemes.

\medskip

\noindent Overall, the results demonstrate that microwave-driven plasma wakefields in rectangular waveguides can sustain controlled electron acceleration in a regime characterized by moderate energy gains and meter-scale interaction lengths. Although the achievable acceleration gradients remain several orders of magnitude lower than those typical of laser-driven wakefield accelerators, the present study provides a consistent and quantitative assessment of the acceleration stage in this alternative parameter regime. These findings contribute to defining the operational limits and physical constraints of microwave-based plasma acceleration and offer guidance for future experimental and numerical investigations aimed at enhancing performance through optimized plasma densities, pulse parameters, and injection strategies.
%============================================================================
\section*{Acknowledgments}
%============================================================================
\noindent The authors thank the Universidad Industrial de Santander (UIS), Colombia, for supporting this work through internal funding (project ID:~3706).
%============================================================================
\section{REFERENCES}
%============================================================================
\nocite{*}
\bibliography{aipsamp}% Produces the bibliography via BibTeX.

@article{malka2012laser,
  title={Laser plasma accelerators},
  author={Malka, Victor},
  journal={Physics of Plasmas},
  volume={19},
  number={5},
  year={2012},
  publisher={AIP Publishing}
}

@book{wangler2008rf,
  title={RF Linear accelerators},
  author={Wangler, Thomas P},
  year={2008},
  publisher={John Wiley \& Sons}
}

@book{birdsall2004plasma,
  title={Plasma Physics via Computer Simulation},
  author={Birdsall, C. K. and Langdon, A. B.},
  year={2004},
  publisher={CRC Press}
}

@article{umeda2003new,
  title={A new charge conservation method in electromagnetic particle-in-cell simulations},
  author={Umeda, T. and Omura, Y. and Tominaga, T. and Matsumoto, H.},
  journal={Computer Physics Communications},
  volume={156},
  number={1},
  pages={73--85},
  year={2003}
}

@article{lapenta2015kinetic,
  title={Kinetic plasma simulation: Particle in cell method},
  author={Lapenta, Giovanni},
  journal={XII Carolus Magnus Summer School on Plasma and Fusion Energy Physics},
  pages={76--85},
  year={2015}
}

@article{hockney1988computer,
  title={Computer simulation using particles},
  author={Hockney, R.W. and Eastwood, J.W.},
  journal={Taylor \& Francis},
  year={1988}
}

@article{esarey2002overview,
  title={Overview of plasma-based accelerator concepts},
  author={Esarey, Eric and Sprangle, Phillip and Krall, Jonathan and Ting, Antonio},
  journal={IEEE Transactions on plasma science},
  volume={24},
  number={2},
  pages={252--288},
  year={2002},
  publisher={IEEE}
}

@article{albert2016applications,
  title={Applications of laser wakefield accelerator-based light sources},
  author={Albert, F{\'e}licie and Thomas, Alec GR},
  journal={Plasma Physics and Controlled Fusion},
  volume={58},
  number={10},
  pages={103001},
  year={2016},
  publisher={IOP Publishing}
}

@article{esarey2009physics,
  title={Physics of laser-driven plasma-based electron accelerators},
  author={Esarey, Eric and Schroeder, Carl B and Leemans, Wim P},
  journal={Reviews of modern physics},
  year={2009},
  volume={81},
  number={3},
  pages={1229--1285},
  publisher={APS}
}

@article{aria2008wakefield,
  title={Wakefield generation in a plasma filled rectangular waveguide},
  author={Aria, Anil Kumar and Malik, Hitendra Kumar},
  journal={The Open Plasma Physics Journal},
  volume={1},
  number={1},
  year={2008}
}

@article{bliokh2017wakefield,
  title={Wakefield in a waveguide},
  author={Bliokh, YP and Leopold, JG and Shafir, G and Shlapakovski, A and Krasik, Ya E},
  journal={Physics of Plasmas},
  volume={24},
  number={6},
  year={2017},
  publisher={AIP Publishing}
}

@article{krasik2019experiments,
  title={Experiments designed to study the non-linear transition of high-power microwaves through plasmas and gases},
  author={Krasik, Yakov E and Leopold, John G and Shafir, Guy and Cao, Yang and Bliokh, Yuri P and Rostov, Vladislav V and Godyak, Valery and Siman-Tov, Meytal and Gad, Raanan and Fisher, Amnon and others},
  journal={Plasma},
  volume={2},
  number={1},
  pages={51--64},
  year={2019},
  publisher={MDPI}
}

@article{cao2019wakefield,
  title={Wakefield excitation by a powerful sub-nanosecond 28.6 GHz microwave pulse propagating in a plasma filled waveguide},
  author={Cao, Y and Bliokh, Y and Leopold, JG and Rostov, V and Slutsker, Ya and Krasik, Ya E},
  journal={Physics of Plasmas},
  volume={26},
  number={2},
  year={2019},
  publisher={AIP Publishing}
}

@article{cao2024direct,
  title={Direct measurement of the wakefield excited by a high-power microwave pulse in plasma},
  author={Cao, Y and Maksimov, V and Haim, A and Leopold, JG and Kostinskiy, A and Bliokh, YP and Hadas, Y and Krasik, Ya E},
  journal={Physics of Plasmas},
  volume={31},
  number={4},
  year={2024},
  publisher={AIP Publishing}
}

@article{pukhov2015particle,
  title={Particle-in-cell codes for plasma-based particle acceleration},
  author={Pukhov, Alexander},
  journal={arXiv preprint arXiv:1510.01071},
  year={2015}
}

@book{sullivan2013electromagnetic,
  title={Electromagnetic simulation using the FDTD method},
  author={Sullivan, Dennis M},
  year={2013},
  publisher={John Wiley \& Sons}
}

@article{courant1967partial,
  title={On the partial difference equations of mathematical physics},
  author={Courant, Richard and Friedrichs, Kurt and Lewy, Hans},
  journal={IBM journal of Research and Development},
  volume={11},
  number={2},
  pages={215--234},
  year={1967},
  publisher={IBM}
}

@article{cakir2019brief,
  title={A Brief Review of Plasma Wakefield Acceleration},
  author={Cakir, Altan and Guzel, Oguz},
  journal={arXiv preprint arXiv:1908.07207},
  year={2019}
}

@article{tajima1979laser,
  title={Laser electron accelerator},
  author={Tajima, Toshiki and Dawson, John M},
  journal={Physical review letters},
  volume={43},
  number={4},
  pages={267},
  year={1979},
  publisher={APS}
}

@article{maity2025coupling,
  title={Coupling and acceleration of externally injected electron beams in laser-driven plasma wakefields},
  author={Maity, Srimanta and Sasorov, Pavel and Molodozhentsev, Alexander},
  journal={Journal of Physics D: Applied Physics},
  volume={58},
  number={14},
  pages={145204},
  year={2025},
  publisher={IOP Publishing}
}

@article{mirzaie2025progress,
  title={Progress on Ultra-relativistic Electron Acceleration from Laser Wakefield Acceleration},
  author={Mirzaie, Mohammad and Kim, Hyung Taek and Kim, Chulmin and Kim, Kyung Taec},
  journal={Current Optics and Photonics},
  volume={9},
  number={6},
  pages={581--597},
  year={2025},
  publisher={Optical Society of Korea}
}

@article{lopez2025particle,
  title={Particle-in-cell simulations of plasma wakefield formation in microwave waveguides},
  author={L{\'o}pez, Jes{\'u}s E and Orozco-Ospino, Eduardo A},
  journal={Physics of Plasmas},
  volume={32},
  number={11},
  year={2025},
  publisher={AIP Publishing}
}

@article{RevModPhys.81.1229,
  title = {Physics of laser-driven plasma-based electron accelerators},
  author = {Esarey, E. and Schroeder, C. B. and Leemans, W. P.},
  journal = {Rev. Mod. Phys.},
  volume = {81},
  issue = {3},
  pages = {1229--1285},
  numpages = {0},
  year = {2009},
  month = {Aug},
  publisher = {American Physical Society},
  doi = {10.1103/RevModPhys.81.1229},
  url = {https://link.aps.org/doi/10.1103/RevModPhys.81.1229}
}
%============================================================================
\end{document}